\documentclass[12pt,reqno]{amsart}

\usepackage[foot]{amsaddr}
\usepackage[latin1]{inputenc}
\usepackage[T1]{fontenc}
\usepackage{mathptmx}
\usepackage{amssymb}
\usepackage{extarrows}
\usepackage{tikz}

\usetikzlibrary{positioning, calc}
\usetikzlibrary{arrows,decorations.markings}

\usepackage{color}

\newcommand{\ket}[1]{\left|#1\right\rangle}

\newcommand{\suchthat}{\;\ifnum\currentgrouptype=16 \middle\fi|\;}

\usepackage{paralist}

\usepackage{mathtools}
\DeclarePairedDelimiterX{\norm}[1]{\lVert}{\rVert}{#1}
\newtheorem{theorem}{Theorem}[section]

\theoremstyle{definition}

\newtheorem{definition}[theorem]{Definition}

\numberwithin{equation}{section}

\newcommand{\ZZ}{\mathbb{Z}}

\begin{document}
\title{Quantum collision finding for homomorphic hash functions}
\author[J.C. Garcia-Escartin]{Juan Carlos Garcia-Escartin $^1$}
\address[1]{Departamento de Teor\'ia de la Se\~{n}al y Comunicaciones e Ingenier\'ia Telem\'atica. ETSI de Telecomunicaci\'on. Universidad de Valladolid. Campus Miguel Delibes. Paseo Bel\'en 15. 47011 Valladolid. Spain.}
\email{juagar@tel.uva.es}

\author[V. Gimeno]{Vicent Gimeno $^2$}
\address[2]{Universitat Jaume I, Campus de Riu Sec, Departament de Matem\`atiques \& Institut Universitari de Matem\`atiques i Aplicacions de Castell\'o--IMAC, 12071, Caste\-ll\'on de la Plana, Spain.}
\email{gimenov@uji.es}

\author[J.J. Moyano-Fern\'andez]{Julio Jos\'e Moyano-Fern\'andez $^2$}
\email{moyano@uji.es}

\date{\today}

\begin{abstract}
Hash functions are a basic cryptographic primitive. Certain hash functions try to prove security against collision and preimage attacks by reductions to known hard problems. These hash functions usually have some additional properties that allow for that reduction. Hash functions which are additive or multiplicative are vulnerable to a quantum attack using the hidden subgroup problem algorithm for quantum computers. Using a quantum oracle to the hash, we can reconstruct the kernel of the hash function, which is enough to find collisions and second preimages. When the hash functions are additive with respect to the group operation in an Abelian group, there is always an efficient implementation of this attack. We present concrete attack examples to provable hash functions, including a preimage attack to $\oplus$-linear hash functions and for certain multiplicative homomorphic hash schemes.
\end{abstract}

\maketitle
\section{Quantum algorithms in cryptography}
Quantum computing offers efficient algorithms that solve problems for which known classical methods are impractical. A prime example is Shor's algorithm for factoring and the discrete logarithm which runs in polynomial time \cite{Sho97}. Many public key cryptographic protocols are based on these two, or closely related, problems. In order to prepare for future quantum computers, there is an active search of quantum resistant cryptographic systems, which are collectively known as post-quantum cryptography \cite{BL17}.

For many other classical cryptographic protocols, known quantum algorithms are of little or no consequence. For brute force search of the key space in symmetric cryptography, Grover's algorithm \cite{Gro97} can only offer a quadratic speedup, which can quickly be solved by doubling the key length. Similarly, for ideal hash functions, quantum computer can only offer modest speedups \cite{CNS17}. 

While these general attacks are limited, there are still quantum attacks that are efficient against particular families of symmetric cryptosystems. For instance, symmetric ciphers based on the Even-Mansour construction become insecure in a quantum setting \cite{KM12} and certain common modes of operation in authentication and authenticated encryption can be attacked with quantum period finding \cite{KLL16}.

In this paper, we show that certain families of cryptographic hash functions that are additive or multiplicative are vulnerable to quantum attacks. These functions are sometimes the basic element in homomorphic hash applications \cite{KFM04}.

\section{Cryptographic hash functions}
An ideal hash function is a function $H(x)=y$ which takes an input binary string of an arbitrary length $x$ into an output $y \in \{0,1\}^n$ with a fixed number of bits $n$. Depending on the intended use, there are many definitions of what constitutes a proper cryptographic hash function. Some common requirements, in a broad formulation, are \cite{MVO96}:
\begin{itemize}
\item \emph{Collision resistance}: It should be infeasible to find two values $x, x'$ with $x\neq x'$ such that $H(x)=H(x')$.
\item \emph{Preimage resistance}: For a fixed hash value $y$ it should be infeasible to find a string $x$ such that $H(x)=y$.
\item \emph{Second
 preimage resistance}: For a fixed input $x$ it should be infeasible to find a second string $x'$ such that $H(x)=H(x')$.
\end{itemize}

In practice, we can consider ideal hash functions as random transformations that take any input $x$ into a random string of $n$ bits and for which even the smallest change in $x$ (1 flipped bit) results in completely new output (which has, on average, only half bits in common with the first hash).

We present a second preimage attack for hash functions that are additive or multiplicative (see Section \ref{homomorphic}). This automatically gives a family of collisions. With a number of operations polynomial in the number of input bits we can find an exhaustive list of collisions for any desired input.

\subsection{Homomorphic hash functions}
\label{homomorphic}
A general hash function works on lengths of an arbitrary ouput. In the following we are adopting a definition with a fixed input size: 
\begin{definition}
An $m$-to-$n$ hash function $H(x): \{0,1\}^m \to \{0,1\}^n$ is a function that takes an $m$-bit string $x$ into an $n$-bit string $y$ with $m>n$. 
\end{definition}
In the following, we will use the term hash function to talk about $m$-to-$n$ hash functions. This covers some existing fixed-size hash functions and the general case, where we have to restrict to inputs of the same size as the string for which we want a collision. In both cases we can obtain a valid collision (or a second preimage).

Hash functions in cryptography should be inversion, collision and preimage resistant. In many cases, this resistance is assumed from the statistical mixing inside the function. However, in the functions generally known as {\it provably secure} hash functions, resistance to attacks is founded on reductions to assumed hard problems (like factoring or the discrete logarithm problem). Proofs are possible because of an additional imposed structure on the functions. Similarly, for some applications like homomorphic encryption, there are additional properties which prevent the concerned hash functions to behave as fully random transformations. This is usually not a problem for many applications as long as we can keep collision resistance or similar properties. 

Many provable hash functions have an additive or multiplicative property, depending on the group operation. These functions are defined by a homomorphism in that group.

\begin{definition}
\label{additive}
A hash function $H(x)$ is {\bf homomorphic} if, for any input pair $x$ and $y$, $H(x+y)=H(x)+H(y)$ for the group operation $+$ in the input and output groups.  
\end{definition}
For instance, $l$-bit strings, together with the XOR operation, form an Abelian group so that a hash function $H(x\oplus y)=H(x)\oplus H(y)$ for the bitwise XOR for $m$ (input) and $n$ bits (output) is an additive hash function.

In the paper we will speak of additive functions and work with groups $(G,+)$. In some contexts, the most natural way to think of the group operation is as a product $\times$ (and to replace the null element by a unit element). Apart from this unimportant nomenclature issue, additive (or multiplicative) hash functions have the same behaviour and are subject to quantum attacks that can help to find collisions.

\section{The hidden subgroup problem and hash collisions}
The most notable quantum algorithms which offer superpolynomial speedups over classical known methods, like Shor's algorithm, solve instances of the hidden subgroup problem \cite{BL95,Lom04,CD10}.

\begin{definition}
Let $G$ be a finite group with a group operation $+$ which can be computed efficiently for any pair $x,y \in G$. Let $f: G \to S$ be a function on the group for some set of values $S$ that defines a subgroup $H=\{k\in G: f(k+g)=f(g) \text{ for all } g \in G \}$. {\bf The hidden subgroup problem} consists in finding a set of generators of this $H<G$ given $f$ and $G$.
\end{definition}

For Abelian groups, quantum computers can solve the hidden subgroup problem efficiently \cite{Lom04}.

The additive hash functions of Definition \ref{additive} take the elements of a Boolean group $G=( \{0,1\}^m,+)$ to the set $S=\{0,1\}^n$ and play the role of the hidding function $f$. For any $x,y \in \{0,1\}^m$, $H(x+y)=H(x)+H(y)$. 

For the collision attack, we consider the hidden subgroup $K$ defined by elements $y$ for which $H(y)$ is the identity element with respect to $+$ in $\{0,1\}^n$. We call this identity element $e$ the null element of the sum and denote it by $\mathbf{0}$. The subgroup $K$ is the kernel of the hash function $H$. 

The additive hash functions we consider are group homomorphisms and their kernel is a normal subgroup of $G$ \cite{Lau03}.

If we can find an element in the subgroup $K=\{y_i \suchthat H(y_i)=\mathbf{0}\}$ we have a preimage attack. $H(x+y_i)=H(x)+H(y_i)=H(x)+\mathbf{0}=H(x)$ and we have two inputs $x+y_i$ and $x$ mapping to the same output. The only exception is the $y_0$ equal to the identity element in the origin group $\mathbf{0}$, which is always an element of $K$. For hash functions, $m>n$ and the order (number of elements) of the input group is always greater that the order of the output set ($|G|>|S|$). The hash function can never be injective and the kernel has at least one element apart from the identity of $G$. 

\subsection{Quantum computers}
In a quantum computer, we will represent each element in a group $G$ with $|G|=M$ by states $\ket{k}$ with a label $k$ for each integer $0<k<M$. When $M=2^m$ we can alternatively write the integer as the corresponding binary string. These states will form a basis for all the possible states $\ket{\psi}=\sum_{k=0}^M \alpha_k \ket{k}$ with complex $\alpha_k $ so that $\sum_{k=0}^M |\alpha_k|^2=1$. For binary strings, we can also write state $\ket{k}$ in terms of $m$ individual qubits $\ket{k}=\ket{k_{m-1}}\cdots \ket{k_{1}}\ket{k_{0}}$. 

All the operations on the state, except for measurement, are reversible and can be written as a unitary $M\times M$ matrix $U$. We use the usual notation $U_1\otimes U_2$ and $U^{\otimes n}$ to denote the tensor product of the operations $U_1$ and $U_2$ and the $U$ operation applied to $n$ different inputs (of the corresponding dimension) respectively.

One particularly useful evolution on a single qubit is given by the Hadamard gate $H\ket{x}=\frac{\ket{0}+(-1)^x \ket{1}}{\sqrt{2}}$. Among other uses, it can be used to prepare uniform superpositions starting from an initial $\ket{0}\cdots\ket{0}\ket{0}$ state. 

While quantum states can be in superpositions of multiple values, in order to retrieve information from the system we need to perform a measurement and we can only recover a single value. Thus the advantage of quantum computing lies not in superpositions alone but in being able to choose a quantum evolution $U$ which results in a destructive interference for the states we are not interested in and a constructive interference between the states we want. For that reason, where quantum computers really shine is in problems with a strong hidden structure where we can extract global properties which are usually inaccessible to classical computers without heavy sampling (checking most of the possible values)

A more detailed description of quantum computing can be found in standard textbooks \cite{NC00,Mer07}.

\subsubsection{Quantum Fourier Transform in finite Abelian groups} 
\label{QFT}
A key operation in quantum computers is the Quantum Fourier Transform, which helps us to produce the necessary constructive and destructive interference which reveals the solution we search for. In this Section, we describe its implementation for Abelian groups. We first need a few definitions.

\begin{definition}
A finite Abelian group $(G,+)$ has $|G|$ distinct one-dimensional irreducible representations called {\bf characters}. A character is a multiplicative function $\chi: G \to \mathbb{C} \setminus  \{0\}$ so that, for the $+$ operation in $G$, $\chi(x+y)=\chi(x)\chi(y)$ for any pair $x,y \in G$.
\end{definition}

From the structure theorem for finite abelian groups, $G$ can be written as a direct sum $G=\ZZ_{N_1}\oplus \cdots \oplus \ZZ_{N_k}$ of $k$ cyclic groups $\ZZ_{i}$ of orders $N_i$. The elements $g \in G$ can be described as $k$-tuples $g=(g_1,\ldots,g_k)$ taking each $g_j$ as an integer modulo $N_j$. The identity of $G$ becomes $e=\mathbf{0}=(0,0,\ldots,0)$.

We can now define a decomposition in terms of each of the cyclic groups from the tuples $\beta_1=(1,0,0,\ldots, 0), \beta_2=(0,1,\ldots, 0),\ldots, \beta_k=(0,0,0,\ldots, 1)$, with all these $\beta_j\in G$ \cite{Lom04}. For any $g=(g_1,g_2,\ldots,g_k) \in G$
\begin{equation}
\chi(g)=\chi\left(\sum_{j=1}^{k}g_j \beta_j\right)=\prod_{j=1}^{k}\chi\left(\beta_j\right)^{g_j}
\end{equation}
where the effect of $\chi$ on any $g$ is completely determined from the values it takes on the $\beta_j$. 

For each $g \in G$, we can define a character $\chi_g(h)=\prod_{j=1}^{k} \omega^{g_j h_j}_{N_j}$ for $h \in G$ and the roots of unity $\omega_{N_j}=e^{i\frac{2\pi}{N_j}}$.

For any fixed character of a finite Abelian group $\chi$:
\begin{equation}
\label{CharSum}
\sum_{g\in G}\chi(g)=\begin{cases}
               |G| \text{ if } \chi=\chi_e\\
               0   \text{ if } \chi\neq\chi_e
            \end{cases}
\end{equation}
where $\chi_e$ is the identitity character which sends any $g \in G$ to 1.

We define a quantum Fourier transform over $G$, $QFT_{G}$ from a character as the operator:
\begin{equation}
QFT_{G}\ket{g}=\frac{1}{\sqrt{|G|}}\sum_{h \in G} \chi_h(g) \ket{h}.
\end{equation}
For a cyclic group $G=\ZZ_N$, the characters $\chi_h(g)=e^{i\frac{2\pi h g }{N}}$ are defined from the roots of unity. We can similarly compute the characters for any group that is a known direct product of cyclic groups. 

We can build efficient quantum circuits giving the $QFT_{G}$ operation for finite Abelian groups. For $G=\ZZ_2^n$, the most common group when working with binary strings, the operation $H^{\otimes n}$ (a Hadamard gate on each qubit) gives an efficient implementation. Similarly, for any cyclic group $G=\ZZ_N$, even for an unknown $N$, the Quantum Fourier Transform
\begin{equation}
QFT_{G}\ket{x}=\sum_{y\in G} \omega_N^{xy}\ket{y}
\end{equation}
with $\omega_N=e^{i\frac{2\pi}{N}}$ can be computed efficiently (and is indeed the QFT used in Shor's algorithm) \cite{HH00,Lom04}.

For a group with a know factorization $G=\ZZ_{U_1}\times \cdots \times\ZZ_{U_{k-1}}\times\ZZ_{U_{k}}$ (using a direct product notation), 
there are also efficient constructions using the unitary evolution $QFT_G=QFT_{U_1}\otimes \cdots \otimes QFT_{U_{k-1}}\otimes QFT_{U_k}$ resulting from the tensor product of the QFT in each known cyclic group.  

In fact, for any Abelian group, we can approximate the corresponding Quantum Fourier Transform and even use a simpler version that still works as expected for the Hidden Subgroup Problem using Fourier Sampling \cite{HH00,CD10}.

The hash functions we review are all defined for finite Abelian groups, but there exist QFT generalizations which could help in additional problems \cite{MRR06,GSV01}.

\subsubsection{Orthogonal subgroups and cosets}

We also used two important results related to any subgroup $H<G$.

\begin{definition}
For a subset $X \subseteq G$, we say $h \in G$ is {\bf orthogonal} to $X$ if $\chi_h(x)=1$ for all $x \in X$. 
\end{definition}
\begin{definition}
\label{Horth}
For any subgroup $X < G$, the {\bf orthogonal subgroup} $H^{\perp}=\{g\in G \suchthat \chi_g(h)=1 \text{ for all } h\in H\}$ is the set of all the elements in $G$ orthogonal to $H$. This $H^{\perp}$ is a subgroup of $G$ and determines $H$ uniquely.
\end{definition}

\begin{definition}
Let $H$ be a subgroup of $(G,+)$. For a fixed element $g_i\in G$, the {\bf left coset} is the set $g_iH=\left\{g_i+h \text{ for all } h \in H\right\}$ and the {\bf right coset} is $Hg_i=\left\{h+g_i \text{ for all } h \in H\right\}$. For an Abelian group both cosets are the same.
\end{definition}

A key result for the Fourier Transform over Abelian groups is that it takes uniform superpositions from a subgroup $H$ into a uniform superposition in the orthogonal subgroup $H^{\perp}$ \cite{Lom04}
\begin{equation}
QFT_G\left( \frac{1}{\sqrt{|H|}}\sum_{h \in H}\ket{h} \right)=\frac{1}{\sqrt{|H^{\perp}|}}\sum_{h' \in H^{\perp}}\ket{h'}.
\end{equation}
\begin{proof}
We have
\begin{eqnarray}
QFT_G\left( \frac{1}{\sqrt{|H|}}\sum_{h \in H}\ket{h}\right) &=& \frac{1}{\sqrt{|H|}}\sum_{h \in H}QFT_G\ket{h}= \frac{1}{\sqrt{|G||H|}}\sum_{h \in H}\sum_{g \in G}\chi_g(h)\ket{g} \\
&=&\frac{1}{\sqrt{|G||H|}}\sum_{g \in G}\left(\sum_{h \in H}\chi_g(h)\right)\ket{g}.
\end{eqnarray}
The character $\chi_g$ of $G$ is also a character of $H$ and the sum is 0 unless it is the identity on $H$, when it becomes $|H|$ (see Eq. (\ref{CharSum})). That $\chi_g(h)=1$ for all the elements $h\in H$ is precisely the definition of the elements of the orthogonal subgroup $H^{\perp}$ (see Definition \ref{Horth}). So
\begin{eqnarray}
\frac{1}{\sqrt{|G||H|}}\sum_{g \in G}\left(\sum_{h \in H}\chi_g(h)\right)\ket{g}&=& \frac{1}{\sqrt{|G||H|}}\sum_{g \in H^{\perp}}|H|\ket{g}= \sqrt{\frac{|H|}{|G|}}\sum_{g \in H^{\perp}}\ket{g},
\end{eqnarray}
which is a uniform superposition over $H^{\perp}$ which is $G/H$ and has $\frac{|G|}{|H|}$ elements.

\end{proof}

Assuming an Abelian group, which is the case for the additive hash functions under study, we call $H_i$ to the coset $g_iH=Hg_i$. We are concerned with the Fourier Transform
\begin{equation}
QFT_G \left(\frac{1}{\sqrt{|K|} }\sum_{g\in H_i} \ket{g}\right)=\frac{1}{\sqrt{|K^{\perp}|}}\sum_{h\in K^{\perp}} \chi_{h}(g_i)\ket{h}
\end{equation}
for any fixed $g_i$ (representative) giving the coset $H_i$.

A quantum collision algorithm will sample random elements from $H_i$ until it can deduce a generating set for $K$. Each element of $H^{\perp}$ gives one condition in a system of equations which completely describes $H$ after sampling a number of orthogonal elements logarithmic with the size of $G$.

\section{General collision algorithm}
The tools from the previous sections allow us to define a general collision finding algorithm with the following steps:

\begin{itemize}
\item Prepare an initial state $\ket{0}\ket{0}$ with two registers, the first with $m$ qubits, the second with $n$.

\item Create a uniform superposition 
\begin{equation}
\frac{1}{\sqrt{M}}\sum_{x=0}^{M-1}\ket{x} \ket{0}.
\end{equation}
This can be done with a $H^{\otimes m}\otimes I^{\otimes n}$ or, depending on our group, $QFT_G \otimes I^{\otimes n}$.

\item Call the hash oracle to transform the uniform superposition into 
\begin{equation}
\frac{1}{\sqrt{M}}\sum_{x=0}^{M-1}\ket{x} \ket{H(x)}.
\end{equation}
For binary strings, we use the usual unitary $U_f \ket{x}\ket{0}=\ket{x}\ket{y\oplus f(x)}$, which can be always implemented for functions $f$ with an efficient classical implementation (as hash functions should). In other groups, such as the multiplicative group of integers modulo $N$, we can use modular addition. In general, for a $+$ operation in the image group of $H$, we have an efficient method to map the null element into $H(x)$.

\item Measure the second register. The new quantum state is 
\begin{equation}
\frac{1}{\sqrt{|K|}}\sum_{y_i\in K}\ket{x_0+y_i} \ket{H(x_0)}.
\end{equation}
We use that $H(x+y)=H(x)+H(y)$ for the $y_i\in K$. For $m>n$ (any useful hash function), there will be more than one value mapping to the same $h$. We call $x_0$ to the smallest such value.

The result is a uniform superposition over the values $x_0+y_i$ for all the $y_i$ in the desired subgroup (the kernel of the hash function $H$). The second register can be ignored from this point. 

\item Compute the $QFT_G$ of the first register in the corresponding Abelian group. The first register has a uniform superposition of the elements in the $x_0K$ coset and the result will be a uniform superposition of the elements of the orthogonal subgroup $K^{\perp}$ with
\begin{equation}
QFT_G\left(\frac{1}{\sqrt{|K|}}\sum_{y_i\in K}\ket{x_0+y_i} \ket{H(x_0)}\right)=\frac{1}{\sqrt{|K^{\perp}|}}\sum_{z\in K^{\perp}}\chi_{z}(x_0)\ket{z}\ket{H(x_0)}.
\end{equation}
Before $QFT_G$, measuring the first register would only give an input/output pair. We exploit the hidden structure to force a destructive intereference for all the elements outside the orthogonal group. 

\item Measure the first register to find a random element of $K^{\perp}$ with equal probability. 
\end{itemize}

This finishes the quantum part. Once we have a random sample of the orthogonal subgroup, we obtain a new restriction to the possible elements in the generating set of $K$. We repeat the process until we have enough information to find the whole generator. Strictly speaking, for a collision or preimage attack, it suffices to find one element $y_k\neq \mathbf{0} \in K$. We can stop as soon as we get the first random element of $K$ which is not the identity. Then, for any input string $x$, we can generate a message $x \oplus y_k$ so that $H(x\oplus y_k)=H(x)+H(y_k)=H(x)$.

The method is efficient as long as:
\begin{itemize}
\item We can efficiently generate a uniform superposition over the group $G$.

Typically, we need access to inputs which are arbitary binary strings (we can restrict to $m$ bits with each attack) or integers in a range from $0$ to $N$ (usually converted from a binary string). In both cases it is easy to create the superposition either from the $\ket{0}$ string and a Hadamard gate for each bit (input bits) or from the $\ket{0}$ state and the $QFT$ as used in Shor's algorithm (integers).

\item We have an efficient quantum function computing $H(x)$ for $x \in G$. The classical hash function must have a reasonable computation time in order to be useful. Any classical binary function can be converted into a reversible function if we keep the input and compute $\ket{x}\ket{y}\to\ket{x}\ket{y\oplus H(x)}$ for a bitwise XOR operation $\oplus$, which is enough to go from $\ket{x}\ket{0}$ to $\ket{x}\ket{H(x)}$.

\item There is an efficient Quantum Fourier Transform. For Abelian groups, we have seen in Section \ref{QFT} there are either efficient quantum circuits or good approximations which can still be used to find elements in the orthogonal subgroup.
\end{itemize}

In particular, for binary strings and the XOR operation, we have simple quantum circuits. The set of binary strings with $m$ bits, together with the bitwise XOR operation, forms an Abelian group which can be written as $\ZZ_2\times \ldots \times\ZZ_2$ with $m$ factors. For this decomposition, in each of the cyclic groups associated to each bit, $\omega_2=e^{i\frac{2\pi}{2}}=-1$ is a root of unity and the character $\chi_{g_i}(h_i)=(-1)^{g_i h_i}$ is a valid character for the possible values $g_i, h_i \in \ZZ_2$ that correspond to $i$th bits of $g$ and $h$. Then, we have a valid character for $m$-bit strings and the XOR operation:
\begin{equation}
\chi_g(h)=\prod_{i=1}^{m} (-1)^{g_i h_i}=(-1)^{g\cdot h}
\end{equation}
where $g\cdot h$ is the inner product on the bit strings representing $g$ and $h$ (the parity of the bitwise AND of the strings). 

For this character, the quantum Fourier transform in the group can be written as
\begin{equation}
QFT_G\ket{g}=\frac{1}{\sqrt{M}}\sum_{h \in G} \chi_g(h)\ket{h}=\frac{1}{\sqrt{M}}\sum_{h \in G}(-1)^{g\cdot h}\ket{h},
\end{equation}
which corresponds to the quantum operation $QFT_G=H^{\otimes m}$ (applying a Hadamard gate to each qubit).

With this Fourier transform we get a random $z$ in the subgroup orthogonal to the kernel, $z\in K^{\perp} \iff (-1)^{x\cdot z}=1$ for all $x\in K$. Any two elements $z \in K^{\perp}$ and $x \in K$ obey $z\cdot x=0$. 

Each measurement gives a restriction to the possible values of the elements in $K^\perp$, which allows us to discover a generating set of $K^\perp$ after a number of measurements polynomial in the number of bits $m$.

Furthermore, once we have a generating set of $K^{\perp}$, we can compute a random element in $K$ efficiently (polynomial time in $m$) and, from that, a generating set of $K$ in expected polynomial time. The classical method is described in \cite{Lom04,Dam04}. Basically, each measurement gives, with high probability, a new equation from a linear system which can be solved to obtain a generating set for $K$. For additive hash functions the system will always have a solution and the result can be used to find collisions or a second preimage to any input $x$. This completes the attack. 

\section{Examples}
In this Section, we examine some hash proposals which would be insecure under our quantum attack. Somewhat ironically, these functions try to guarantee security against collisions by reduction to a hard problem, but the additional structure imposed on the functions allow for a quantum attack.

\subsection{$\oplus$-linear hash functions}
In \cite{Kra94} Krawczyk presented two families of $\oplus$-linear hash functions $H(x): \{0,1\}^m\to\{0,1\}^n$ which are additive with respect to the XOR operation. For any two inputs $x_1, x_2 \in \{0,1\}^m$, $H(x_1\oplus x_2)=H(x_1)\oplus H(x_2)$. 

The designs are based on Cyclic Redundancy Codes and Linear Feedback Shift Registers and have some some desirable properties. For instance, uniformity can be proved instead of assumed like in most hash functions. Unfortunately, the $\oplus$-linearity also opens the door for a quantum attack.

The attack is, basically, a quantum algorithm for a generalized Simon's problem. In the original Simon's algorithm we have a promise that a function $f(x): \{0,1\}^n\to \{0,1\}^n$ such that $f(x\oplus s)=f(x)$ only for two values (with a secret string $s$). Here we have a slightly different problem. For a balanced function there will be $2^{m-n}$ strings with the same output value. The group is $(\{0,1\}^m,\oplus)$ and the hidden subgroup is the kernel of $H(x)$. After the quantum algorithm we get elements $y_i$ with $H(y_i)=\mathbf{0}$ so that $H(x\oplus y_i)=H(x)\oplus H(y_i)=H(x)$.

\subsection{Homomorphic hash function with multiplication}
The attack can be translated to multiplicative hashes in groups where the group operation is more naturally cast as a multiplication and the null element as the unit. 

We are going to see two examples with hashes in the multiplicative group of integers modulo $N$ (the group of units in $\ZZ_N$). The group operation is multiplication modulo $N$ and the identity element is the integer 1.

Our first example is the RSA hash $E(x)=x^e \mod N$ for an $N=pq$ with unknown factorization, which has a multiplicative property: $E(xy)=E(x)E(y)$. The proposed attack finds the kernel consisting in all the messages $x_i$ for which $E(x_i)=1$. This particular example is not useful as a hash function. It depends on trusting no one knows the factorization of $N$. 

However, multiplicative and additive properties appear in many proposals for homomorphic encryption and any hash function derived from them should be checked against quantum attacks.

For instance, the collision resistant hash function used in the homomorphic hash scheme proposed by Krohn, Freedman and Mazi\`eres \cite{KFM04} is vulnerable to a quantum attack. The basic transformation is defined as
\begin{equation}
h_G({\bf b}_j)=\prod_{i=1}^{m}g_i^{b_{i,j}}\mod p
\end{equation}
for a message block ${\bf b}_j$ composed of $m$ integers $b_{i,j}$ from 0 to a prime $q$ dividing $p-1$. The integer $p$ is a random prime and $g_i$ are randomly chosen integers of order $q$ modulo $p$. For any two blocks ${\bf b}_i$ and ${\bf b}_j$,
\begin{equation} 
h_G({\bf b}_i+{\bf b}_j)=h_G({\bf b}_i)h_G({\bf b}_j),
\end{equation}
where ${\bf b}_i+{\bf b}_j$ is a vector with elements $b_{1,i}+b_{1,j} \mod q$ to $b_{m,i}+b_{m,j} \mod q$.

The inputs are vectors with elements in the additive group of integers modulo $q$ and the hash takes them into the group of units modulo $\ZZ_p$. Finding a kernel for $h_G$ gives blocks ${\bf b}_e$ with $h_G({\bf b}_e)=1$, which yield collisions for any desired input block ${\bf b}_i$. 

The hash function is a compression function and the kernel will contain multiple elements. Most of them will be useful for collisions with two exceptions. First, the kernel will always contain a trivial zero block which maps each block to itself and for which all the $b_{i,j}$ are 0. Second, some of the blocks ${\bf b}_e$ might not correspond to valid binary sequences. The number of binary digits $n$ for each block is chosen so that $2^n<q$ and, if any of the $b_{i,e}$ is greater than $2^n$, there is no binary input corresponding to that integer. In the attack we can always fix some of the blocks of the input to 0 so that this happens with an acceptably small probability.

\section{Discussion}
We have shown quantum computers can find collisions for additive hash functions by finding its kernel subgroup. The attack is valid for hash functions with a strong structure, such as those usually proposed for provably secure hashing.

We have given examples of the attack working on the $\oplus$-linear hash functions of Krawczyk \cite{Kra94} and on certain homomorphic hashing schemes \cite{KFM04}.

As opposed to some previous quantum decryption algorithms, which should have access to a quantum oracle encrypting with an unknown key, the attacker can always find a quantum version of the function and produce the required superpositions. 

Like all collision attacks, quantum collision finding can be performed offline using any fixed input $x_0$. Once the kernel of $H(x)$ has been found, it can be directly used for second preimage attacks in real time to find $H(x')=H(x)$ by adding to the known input $x$ any linear combination of the $y_i$ in the kernel. 

The kernel can also help to craft fake messages that replace a signed string. For instance, for the group of binary strings of $n$ bits under the XOR operation, the attacker can try to alter specific bits from a message by XORing the input string with elements from the kernel that change only the target part of the message and, maybe, also unimportant bits which will not be noticed (like color or gray level bits in a picture). It is not obvious how to perform this kind of attack and it would be highly dependent on the particular structure of the kernel and the concrete addition operation of the relevant group, but it could reduce the complexity of a forgery, at least for some specific scenarios.

The attack exposes a general problem of hash functions: either there is a formal proof of security at the cost of imposing a structure or we are limited to transformations which appear to be random but are difficult to analyze. 

In that respect, many provable hash functions use reductions to problems which can be solved efficiently on a quantum computer, such as factoring, and could be vulnerable to quantum attacks. The quantum security of these hash functions should be studied further. The attacks might not be straightforward. In many cases the reduction is not shown in both directions: finding a collision might solve factoring but it is not known whether factoring provides a collision or not. 

It is also open whether the collision finding method of this paper can be extended or not to other provable hash functions with more complex additive or multiplicative properties. Some possible candidates are VHS \cite{CLS06}, where $H(\mathbf{0})H(x \lor y)\equiv H(x)H(y) \mod N$ for $x\land y=\mathbf{0}$, or the muHASH, adHASH and LtHASH families \cite{BM97}.

\section*{Acknowledgements}
The first author has been funded by the Spanish Government and FEDER grant PID2020-119418GB-I00 (MICINN) and by Junta de Castilla y Le\'on (project VA296P18). The second author has been partially supported by the Research Program of the University Jaume I--Project UJI-B2018-35, as well as by the Spanish Government and FEDER grant PID2020-115930GA-I00 (MICINN). The third author was partially supported by the Spanish Government, Ministerios de Ciencia e Innovaci\'on y de Universidades, grant PGC2018-096446-B-C22, as well as by Universitat Jaume I, grant UJI-B2018-10.

\bibliographystyle{amsalpha} 
\newcommand{\noopsort}[1]{} \newcommand{\printfirst}[2]{#1}
  \newcommand{\singleletter}[1]{#1} \newcommand{\switchargs}[2]{#2#1}
\providecommand{\bysame}{\leavevmode\hbox to3em{\hrulefill}\thinspace}
\providecommand{\MR}{\relax\ifhmode\unskip\space\fi MR }
\providecommand{\MRhref}[2]{%
  \href{http://www.ams.org/mathscinet-getitem?mr=#1}{#2}
}
\providecommand{\href}[2]{#2}
 
\end{document}